\documentclass[twocolumn,aps,english,prl,nofootinbib,preprintnumbers]{revtex4}
\usepackage{mathrsfs}

\usepackage{mathrsfs}
\usepackage{graphicx}
\usepackage{amsmath, amssymb}
\usepackage{babel}
\usepackage{color}
\usepackage{slashed}
\usepackage{hyperref}
\usepackage[T1]{fontenc}
\usepackage[export]{adjustbox}

\begin{document}
	
	\title{Electroweak nuclear radii constrain the isospin breaking correction to $V_{ud}$}

	\author{Chien-Yeah Seng$^{1,2,3}$}
	\author{Mikhail Gorchtein$^{4,5}$}

	\affiliation{$^{1}$Helmholtz-Institut f\"{u}r Strahlen- und Kernphysik and Bethe Center for
		Theoretical Physics,\\ Universit\"{a}t Bonn, 53115 Bonn, Germany}
	\affiliation{$^{2}$Facility for Rare Isotope Beams, Michigan State University, East Lansing, MI 48824, USA}
	\affiliation{$^{3}$Department of Physics, University of Washington,
		Seattle, WA 98195-1560, USA}
	\affiliation{$^{4}$Institut f\"ur Kernphysik, Johannes Gutenberg-Universit\"{a}t,\\
		J.J. Becher-Weg 45, 55128 Mainz, Germany}
	\affiliation{$^{5}$PRISMA Cluster of Excellence, Johannes Gutenberg-Universit\"{a}t, Mainz, Germany}

	\date{\today}
	
\begin{abstract}
		
We lay out a novel formalism to connect the isospin-symmetry breaking correction to the rates of superallowed nuclear beta decays, $\delta_\text{C}$, to the isospin-breaking sensitive combinations of electroweak nuclear radii that can be accessed experimentally. We individuate transitions in the superallowed decay chart where a measurement of the neutron skin of a stable daughter even
at a moderate precision could already help discriminating between models used to compute $\delta_\text{C}$. We review the existing experimental situation and make connection to
the existing and future experimental programs.
		
\end{abstract}
	
\maketitle
	
{\bf\boldmath Introduction} -- Superallowed beta decays of $0^+$ nuclei provide currently the best measurement of the
Cabibbo-Kobayashi-Maskawa (CKM) matrix element $V_{ud}$, which permits high-precision tests of the Standard Model (SM) prediction through the first-row unitarity constraint $|V_{ud}|^2+|V_{us}|^2+|V_{ud}|^2=1$. Recent improvements in the single-nucleon radiative correction theory~\cite{Seng:2018yzq,Seng:2018qru,Czarnecki:2019mwq,Seng:2020wjq,Hayen:2020cxh,Shiells:2020fqp} unveil an apparent violation of the unitarity relation~\cite{ParticleDataGroup:2022pth} and motivate the renewed interest within the theory and experimental communities~\cite{Crivellin:2020lzu,Crivellin:2021njn}. The current most precise extraction of $|V_{ud}|$ is obtained from a global analysis of superallowed nuclear decays following the master formula~\cite{Hardy:2020qwl}
\begin{align}
	|V_{ud}|^2\,\mathcal{F}t \,(1+\Delta_R^V)=2984.43\,\mathrm{s}\,.
	\label{eq:VudExtr}
\end{align}
Among the ingredients in the above formula, the r.h.s. combines the very precisely known physical constants with uncertainties far beyond the precision goal relevant to the analysis of beta decays. The l.h.s., along with $|V_{ud}|^2$ and the already mentioned single-nucleon radiative correction $\Delta_R^V$, contains the universal, decay-independent $\mathcal{F}t$-value. The latter is defined by absorbing the experimental, process-specific measurements summarized as $ft$, with decay-specific nuclear corrections~\cite{Hardy:2020qwl}, 
\begin{equation}
	\mathcal{F}t=ft(1
	+\delta_R^\prime)(1+\delta_\text{NS})(1-\delta_\text{C}).
	\label{eq:FtDef}
\end{equation}
The QED correction $\delta_R^\prime$ describes soft-photon effects beyond Coulomb distortion, that bear dependence on bulk nuclear properties. 
The nucleon structure correction $\delta_\text{NS}$ encodes the nuclear dependence of the $\gamma W$-box that has to be accounted for upon extracting the single-nucleon radiative correction $\Delta_R^V$ in Eq.~(\ref{eq:VudExtr}). This separation has recently been addressed in Refs.~\cite{Seng:2018qru,Gorchtein:2018fxl}.
	
The isospin-symmetry breaking (ISB) correction $\delta_\text{C}$  modifies the squared Fermi matrix element $M_F$ from its isospin-limit value $M_F^0$ as $|M_F|^2=|M_F^0|^2(1-\delta_\text{C})$ even in the absence of radiative corrections. 
It arises from the isospin mixing of the nuclear states arising predominantly from Coulomb repulsion between the protons in the nucleus. Across the nuclear decays relevant for the high-precision extraction of $V_{ud}$, $\delta_\text{C}$ ranges from $\sim0.1\%$ for the $^{10}$C decay to $\sim1.5\%$ for the $^{74}$Rb decay. 
Thus, it plays a central role in aligning the experimental, nucleus-dependent $ft$-values to a nucleus-independent constant  $\mathcal{F}t$-value, as required by conservation of vector current (CVC)~\cite{Hardy:2020qwl}. 
	
At present, this correction is obtained solely from nuclear model calculations; the nuclear shell model calculations with the Woods-Saxon potential~\cite{Towner:2002rg,Towner:2007np,Hardy:2008gy,Hardy:2014qxa,Hardy:2020qwl} result in an impressive alignment of the $\mathcal{F}t$-values. However, concerns  about possible theory inconsistencies of these calculations~\cite{Miller:2008my,Miller:2009cg,Condren:2022dji}, and significant model dependence~\cite{Satula:2011br,Satula:2016hbs,Ormand:1989hm,Ormand:1995df,Liang:2009pf,Auerbach:2008ut,Damgaard:1969yyx} persist. Modern ab-initio calculations that could help reducing the model dependence are still in the preliminary stage~\cite{Caurier:2002hb,Martin:2021bud}.
With this ongoing discussion on $\delta_\text{C}$ in the nuclear theory community, no direct experimental constraints on the ISB correction exist to our knowledge.
	
In this Letter we explore the connection between
$\delta_\text{C}$ and a set of experimentally accessible quantities that are sensitive to the same ISB nuclear matrix elements. 
These observables encompass recoil effects in the superallowed decay process, nuclear charge radii across the isotriplet, and the neutron skin of the stable daughter nucleus.
The relevant combinations are constructed such that non-ISB contributions cancel out, and a clean probe of the isospin mixing effects is obtained. 	
	
{\bf\boldmath Basic notation} -- We adopt the ``nuclear physics convention" for the isospin projection,  $(T_{z})_p=-1/2$. 
We consider $\beta^+$ transitions $i\rightarrow f$ accross the isotriplet with $T_{z,i}=0$ and $T_{z,f}=+1$ (which we will explain later). The Fermi matrix element is defined as
$M_{F}=\langle f|\hat{\tau}_{+}|i\rangle$,
with $\hat{\tau}_{+}$ the isospin-raising operator, and the states $|i\rangle$,
$|f\rangle$ normalized to 1. 

The nuclear states are eigenstates of the full Hamiltonian $H$ which we split as $H=H_{0}+V$,
with $H_{0}$ the part that conserves isospin and $V$ the ISB perturbation term. We label the eigenstates of $H_{0}$
as $|a;T,T_{z}\rangle$ where $a$ denotes all quantum numbers
unrelated to isospin (we use $a=g$ for the ground state isotriplet that undergoes superallowed beta decay). The corresponding energy eigenvalues are labelled
as $E_{a,T}$, which may depend on $a$ and $T$ but not $T_{z}$. 
In the absence of $V$, the bare Fermi matrix element reads $M_F^0=
\langle g;1,T_{z,f}|\hat{\tau}_+|g;1,T_{z,i}\rangle=\sqrt{2}$.

A key ingredient in our analysis is the isovector monopole operator,
\begin{equation}
\vec{M}^{(1)}=\sum_{i=1}^{A}r_{i}^{2}\vec{\hat{T}}(i)
\end{equation}
where $\vec{\hat{T}}(i)$ is the isospin operator of the nucleon $i$, and $\vec{r}_i$ its position.
The irreducible tensors of rank 1 in the isospin space with its components are: $M_{0}^{(1)}=M_{z}^{(1)}$, $M_{\pm 1}^{(1)}=\mp(M_{x}^{(1)}\pm iM_{y}^{(1)})/\sqrt{2}$.

{\bf\boldmath Key experimental observables} -- The charged weak form factors in superallowed decays of spinless nuclei are:
\begin{equation}
\langle f(p_{f})|J_{W}^{\lambda\dagger}(0)|i(p_{i})\rangle=f_{+}(t)(p_{i}+p_{f})^{\lambda}+f_{-}(t)(p_{i}-p_{f})^{\lambda},
\end{equation}
where $J_{W}^{\lambda\dagger}(x)=\bar{d}(x)\gamma^{\lambda}(1-\gamma_{5})u(x)$
is the charged weak current, 
and $t=(p_i-p_f)^2$. The contribution of $f_{-}(t)$ to the differential decay rate is suppressed simultaneously
by kinematics and by ISB, so we can only probe $f_{+}(t)$. In the Breit frame ($p_i^0=p_f^0$), $f_{+}(0)=M_{F}$ 
and we define
$f_{+}(t)=M_{F}\bar{f}_{+}(t)$ 
with $\bar{f}_{+}(0)=1$. For small $t$ we have,
	\begin{equation}
	\bar{f}_{+}(t)=1+\frac{t}{6}R_\text{CW}^2+\mathcal{O}(t^{2}),\label{eq:fbart}
	\end{equation}
	where
	\begin{equation}
	R_\text{CW}^2\equiv-\frac{\sqrt{2}}{M_F}\langle f|M_{+1}^{(1)}|i\rangle
	\end{equation}
	defines a ``charged weak radius'' associated to the charged weak form factor, and one may safely set $M_{F}\rightarrow\sqrt{2}$ above given our precision goal. This radius
	may in principle be measured through recoil effects in beta decays or neutrino-nucleus scattering. We discuss the feasibility of such measurements in later paragraphs.

Further, we define the root mean square (RMS) radii of the proton and neutron distribution in a nucleus $\phi$ (with the proton number $Z_\phi$ and the neutron number $N_\phi$) as
\begin{equation}
R_{p/n,\phi}=\sqrt{\frac{1}{X}\langle \phi|\sum_{i=1}^{A}r_{i}^{2}\left(\frac{1}{2}\mp\hat{T}_{z}(i)\right)|\phi\rangle},\label{eq:Rpn}
\end{equation}
with $-$ for the proton and $+$ for the neutron and $X=Z_\phi$ or $N_\phi$, respectively. 
These radii naturally connect to the $z$-component of the isovector monopole operator,
\begin{equation}
\langle \phi|M_{0}^{(1)}|\phi\rangle
=\frac{N_\phi}{2}R_{n,\phi}^{2}-\frac{Z_\phi}{2}R_{p,\phi}^{2}.
\end{equation}
In absence of ISB, 
the Wigner-Eckart theorem requires the equality
$\langle g;1,1|M_{+1}^{(1)}|g;1,0\rangle=-\langle g;1,1|M_{0}^{(1)}|g;1,1\rangle$. Hence, the following combined experimental observable
\begin{equation}
\Delta M^{(1)}_A\equiv\langle f|M_{+1}^{(1)}|i\rangle+\langle f|M_{0}^{(1)}|f\rangle
\label{eq:deltaMA}
\end{equation}
offers a very clean probe of ISB effect.
Furthermore, we define another experimentally accessible quantity,
\begin{equation}
\Delta M_B^{(1)}\equiv \frac{1}{2}\left(Z_1 R_{p,1}^2+Z_{-1}R_{p,-1}^2\right)-Z_0 R_{p,0}^2\label{eq:DeltaMB1}
\end{equation}
which combines the $R_p$ across the isotripet ($-1,0,1$ denote $T_z$ of the nucleus). Again, $\Delta M_B^{(1)}$ vanishes in the isospin limit, providing another clean probe of isospin mixing effects. $\Delta M_{A,B}^{(1)}$ are the two key experimental observables that we focus on in this Letter.

While the RMS radii $R_{p,n}$ are generally not observable, they are directly related to nuclear charge and neutral weak radii $R_{\mathrm{Ch},\phi},\,R_{\mathrm{NW},\phi}$. The former are measurable for both stable and unstable nuclear isotopes, mainly from the atomic spectroscopy~\cite{Angeli:2013epw}. The nuclear RMS charge radii are largely given by $R_p$, as the corrections due to the charge radii of the proton and the neutron can easily be included, along with the spin-orbit interaction effects~\cite{Friar:1997js,Sanchez:2006zz,Ong:2010gf,Reinhard:2021gym}.
New results for charge radii of unstable isotopes are anticipated, e.g., from the BECOLA facility at FRIB~\cite{MINAMISONO201385}.

Nuclear weak radii are accessible with  parity-violating electron scattering (PVES) on nuclear targets. The object of interest is the neutron skin $R_{n}-R_{p}\propto R_{\mathrm{NW}}-R_{\mathrm{Ch}}$ which is the subject of a vibrant experimental program at electron scattering facilities~\cite{Kumar:2020ejz,Abrahamyan:2012gp,PREX:2021umo,CREX:2022kgg,Becker:2018ggl} with the scope of obtaining insights into the properties of the neutron-rich matter with relevance for astrophysics~\cite{Reed:2021nqk}. Since fixed-target PVES is only viable with a stable target nucleus, we concentrate on (observationally) stable superallowed daughter nuclei, most of which are  $T_{z,f}=+1$ members of the isotriplet, which motivates the definition of Eq.~(\ref{eq:deltaMA}).
In addition, RMS charge radii of stable nuclei are known to $0.1-0.01\%$ precision~\cite{Angeli:2013epw}, which opens the possibility to extract the respective weak RMS radii with a sub-percent precision ~\cite{Koshchii:2020qkr}.  

The difference in the proton and neutron distributions within a nucleus can generically come from two sources: the neutron excess and ISB effects.  
In asymmetric nuclei with $N>Z$ the skin is mainly generated by the symmetry energy~\cite{Brown:2000pd}, although even there the ISB effects may be non-negligible~\cite{Roca-Maza:2018bpv}. For nearly  symmetric nuclei with $N\approx Z$, such as those participating in the superallowed decays, the ISB effects become comparable. Discussions about the relation between ISB effects and the neutron skin exist in the literature~\cite{Auerbach:2010nz}, but to the best of our knowledge, this is the first time the neutron skin of the  members of a superallowed isotriplet is directly related to 
$\delta_\text{C}$ in that isotriplet.



{\bf\boldmath The connection between $\Delta M^{(1)}_{A,B}$ and $\delta_\textbf{C}$} -- To investigate the underlying physics of $\Delta M^{(1)}_{A,B}$, we resort to the perturbation theory formalism outlined in Refs.\cite{Miller:2008my,Miller:2009cg} The only simplifying assumption is that the ISB operator $V$ predominantly transforms as an isovector
($T=1$, $T_{z}=0$)~\cite{Bertsch:1972xy}. The neglect of the isotensor ISB is likely to introduce an uncertainty of the order of 10-15\%. Inserting the full set of intermediate (isospin-symmetric!) nuclear states, we obtain, 
\begin{eqnarray}
&&\Delta M^{(1)}_A = -\frac{1}{3}\sum_a \frac{\langle a;0||M^{(1)}||g;1\rangle^{*}\langle a;0||V||g;1\rangle}{E_{a,0}-E_{g,1}}\nonumber\\
&& -\frac{1}{2}\sum_{a\neq g}\frac{\langle a;1||M^{(1)}||g;1\rangle^{*}\langle a;1||V||g;1\rangle}{E_{a,1}-E_{g,1}}\nonumber\\
&&-\frac{1}{6}\sum_{a}\frac{\langle a;2||M^{(1)}||g;1\rangle^{*}\langle a;2||V||g;1\rangle}{E_{a,2}-E_{g,1}}\nonumber\\
&&-\sum_{a}\frac{\langle a;2||V||g;1\rangle^{*}\langle a;2||M^{(1)}||g;1\rangle}{E_{a,2}-E_{g,1}} +\mathcal{O}(V^2)\label{eq:DeltaM1Apert}
\end{eqnarray}
and 
\begin{eqnarray}
&&\Delta M^{(1)}_B = \mathfrak{Re}\left\{-\frac{2}{3}\sum_a \frac{\langle a;0||M^{(1)}||g;1\rangle^{*}\langle a;0||V||g;1\rangle}{E_{a,0}-E_{g,1}}\right.\nonumber\\
&& +\sum_{a\neq g}\frac{\langle a;1||M^{(1)}||g;1\rangle^{*}\langle a;1||V||g;1\rangle}{E_{a,1}-E_{g,1}}\nonumber\\
&&\left.-\frac{1}{3}\sum_{a}\frac{\langle a;2||M^{(1)}||g;1\rangle^{*}\langle a;2||V||g;1\rangle}{E_{a,2}-E_{g,1}}\right\} +\mathcal{O}(V^2)\label{eq:DeltaM1Bpert}
\end{eqnarray}
where 
the reduced matrix elements are defined via the Wigner-Eckart
theorem:
\begin{eqnarray}
\langle a;T',T_{z}'|M_{T_{z}''}^{(1)}|g;1,T_{z}\rangle & = & C_{1T_{z};1T_{z}''}^{11;T'T_{z}'}\langle a;T'||M^{(1)}||g;1\rangle\nonumber\\
\langle a;T',T_{z}'|V|g;1,T_{z}\rangle & = & C_{1T_{z};10}^{11;T'T_{z}'}\langle a;T'||V||g;1\rangle,
\end{eqnarray}
with $C$s the Clebsch-Gordan coefficients. Note that our definition of $\Delta M_B^{(1)}$ ensures that the isoscalar operator $\sum_i r_i^2$ in Eq.\eqref{eq:Rpn} does not enter the matrix elements at $\mathcal{O}(V)$. 
Meanwhile, the ISB correction $\delta_{\text{C}}$  starts at $\mathcal{O}(V^{2})$ in accord with the (generalized) Behrends-Sirlin-Ademollo-Gatto theorem~\cite{Behrends:1960nf,Ademollo:1964sr}, and reads
\begin{eqnarray}
&&\delta_{\text{C}}=\frac{1}{3}\sum_{a}\frac{|\langle a;0||V||g;1\rangle|^{2}}{(E_{a,0}-E_{g,1})^{2}}+\frac{1}{2}\sum_{a\neq g}\frac{|\langle a;1||V||g;1\rangle|^{2}}{(E_{a,1}-E_{g,1})^{2}}\nonumber\\
&&-\frac{5}{6}\sum_{a}\frac{|\langle a;2||V||g;1\rangle|^{2}}{(E_{a,2}-E_{g,1})^{2}}+\mathcal{O}(V^{3}).\label{eq:DeltaC}
\end{eqnarray}

Further insight can be obtained with a more detailed information on $V$. It is well known that the dominant
source of the isospin mixing in the nuclear states is played by Coulomb repulsion between protons ~\cite{Radicati:1952zz,MacDonald:1955zza}, with its prevailing part coming from a one-body potential
where each proton is subject to a mean field. Furthermore, we take
the potential of a uniformly charged sphere of radius $R_C$, inside which the whole nucleus resides~\cite{Auerbach:2008ut}:
\begin{equation}
V_{C}\approx-\frac{Ze^{2}}{4\pi R_C^{3}}\sum_{i=1}^{A}\left(\frac{1}{2}r_{i}^{2}-\frac{3}{2}R_C^{2}\right)\left(\frac{1}{2}-\hat{T}_{z}(i)\right).
\end{equation}
While there is an ambiguity that $Z$ is different
across the isotriplet, it is safe to take $Z\approx A/2$, since 
$|T_{z}|\ll Z$. As already mentioned, we disregard the isotensor contributions. In this case, only the isovector component breaks isospin symmetry; taking furthermore into account the fact that the $T_z$ is always a good quantum number as it counts the neutrons and protons in the nucleus, we connect the ISB Coulomb potential with the isovector monopole operator,
\begin{eqnarray}
V_{C}^{(1)} & = & 
(Ze^{2}/8\pi R_C^{3}) M_{0}^{(1)},
\end{eqnarray}
and in what follows we will take
$V=V_{C}^{(1)}$. 
Consequently, we can rewrite Eqs.\eqref{eq:DeltaM1Apert}, \eqref{eq:DeltaM1Bpert} as:
\begin{eqnarray}
\Delta M_A^{(1)}&=&\frac{1}{3}\Gamma_0+\frac{1}{2}\Gamma_1+\frac{7}{6}\Gamma_2+\mathcal{O}(V^2)\nonumber\\
\Delta M_B^{(1)}&=&\frac{2}{3}\Gamma_0-\Gamma_1+\frac{1}{3}\Gamma_2+\mathcal{O}(V^2),\label{eq:DeltaM1pert2}~,
\end{eqnarray}
where
\begin{equation}
\Gamma_T\equiv-\frac{8\pi R_C^3}{Ze^2}\sum_a\frac{|\langle a;T||V_C^{(1)}||g;1\rangle|^2}{E_{a,T}-E_{g,1}}~,
\end{equation}
with $a\neq g$ for $T=1$. This should be compared to the expression for $\delta_\text{C}$ in Eq.\eqref{eq:DeltaC} (with $V\to V_C^{(1)}$). We observe that $\Delta M^{(1)}_{A,B}$ and $\delta_{\text{C}}$ share the same set of reduced matrix elements in the $T=0,1,2$ channels, imposing
a strong experimental constraint on $\delta_{\text{C}}$. This is one of the central results of this work.

The fact that these quantities essentially probe the same underlying physics means that any nuclear theory approach capable to compute $\delta_{\text{C}}$ can also be used to compute $\Delta M^{(1)}_{A,B}$, and thus compared to the experiment. 

\begin{table*}[t]
	\begin{centering}
		\begin{tabular}{|c|c|c|c|c|c|c|c|c|c|c|c|c|c|c|c|}
			\hline 
			Transitions & \multicolumn{1}{c}{} & \multicolumn{1}{c}{} & \multicolumn{1}{c}{$\delta_{\text{C}}$} & \multicolumn{1}{c}{(\%)} &  & \multicolumn{1}{c}{} & \multicolumn{1}{c}{} & \multicolumn{1}{c}{$\Delta M^{(1)}_A$} & \multicolumn{1}{c}{(fm$^{2}$)} &  & \multicolumn{1}{c}{} & \multicolumn{1}{c}{} & \multicolumn{1}{c}{$\left|\frac{\Delta M^{(1)}_A}{AR^{2}/4}\right|$} & \multicolumn{1}{c}{(\%)} & \tabularnewline
			\cline{2-16} 
			& WS & DFT & HF & RPA & Micro & WS & DFT & HF & RPA & Micro & WS & DFT & HF & RPA & Micro\tabularnewline
			\hline 
			\hline 
			$^{26m}\text{Al}\rightarrow^{26}\!\!\text{Mg}$ & 0.310 & 0.329 & 0.30 & 0.139 & 0.08 & -2.2 & -2.3 & -2.1 & -1.0 & -0.6 & 3.2 & 3.3 & 3.0 & 1.4 & 0.8\tabularnewline
			\hline 
			$^{34}\text{Cl}\rightarrow^{34}\!\!\text{S}$ & 0.613 & 0.75 & 0.57 & 0.234 & 0.13 & -5.0 & -6.1 & -4.6 & -1.9 & -1.0 & 4.6 & 5.6 & 4.3 & 1.8 & 1.0\tabularnewline
			\hline 
			$^{38m}\text{K}\rightarrow^{38}\!\!\text{Ar}$ & 0.628 & 1.7 & 0.59 & 0.278 & 0.15 & -5.4 & -14.6 & -5.1 & -2.4 & -1.3 & 4.2 & 11.2 & 3.9 & 1.8 & 1.0\tabularnewline
			\hline 
			$^{42}\text{Sc}\rightarrow^{42}\!\!\text{Ca}$ & 0.690 & 0.77 & 0.42 & 0.333 & 0.18 & -6.2 & -6.9 & -3.8 & -3.0 & -1.6 & 4.0 & 4.5 & 2.5 & 2.0 & 1.1\tabularnewline
			\hline 
			$^{46}\text{V}\rightarrow^{46}\!\!\text{Ti}$ & 0.620 & 0.563 & 0.38 & / & 0.21 & -5.8 & -5.3 & -3.6 & / & -2.0 & 3.3 & 3.0 & 2.0 & / & 1.1\tabularnewline
			\hline 
			$^{50}\text{Mn}\rightarrow^{50}\!\!\text{Cr}$ & 0.660 & 0.476 & 0.35 & / & 0.24 & -6.4 & -4.6 & -3.4 & / & -2.4 & 3.1 & 2.3 & 1.7 & / & 1.2\tabularnewline
			\hline 
			$^{54}\text{Co}\rightarrow^{54}\!\!\text{Fe}$ & 0.770 & 0.586 & 0.44 & 0.319 & 0.28 & -7.8 & -5.9 & -4.4 & -3.2 & -2.8 & 3.3 & 2.5 & 1.9 & 1.4 & 1.2\tabularnewline
			\hline 
		\end{tabular}
		\par\end{centering}
	\caption{\label{tab:final}Estimation of $\Delta M^{(1)}_A$ and $|\Delta M^{(1)}_A/(AR^2/4)|$ from different models. See paragraphs after Eq.\eqref{eq:parameters} for explanations. A few remarks: $A=46,50$ are missing in the RPA calculation, while the DFT calculation gives an unusually large $\delta_\text{C}$ for $A=38$.}
\end{table*}

\begin{table*}[t]
	\begin{centering}
		\begin{tabular}{|c|c|c|c|c|c|c|c|c|c|c|}
			\hline 
			Transitions & \multicolumn{1}{c}{} & \multicolumn{1}{c}{} & \multicolumn{1}{c}{$\Delta M_{B}^{(1)}$} & \multicolumn{1}{c}{(fm$^{2}$)} &  & \multicolumn{1}{c}{} & \multicolumn{1}{c}{} & \multicolumn{1}{c}{$\left|\frac{\Delta M_{B}^{(1)}}{AR^{2}/2}\right|$} & \multicolumn{1}{c}{(\%)} & \tabularnewline
			\cline{2-11} 
			& WS & DFT & HF & RPA & Micro & WS & DFT & HF & RPA & Micro\tabularnewline
			\hline 
			\hline 
			$^{26m}\text{Al}\rightarrow^{26}\!\!\text{Mg}$ & -0.12 & -0.12 & -0.11 & -0.05 & -0.03 & 0.08 & 0.09 & 0.08 & 0.04 & 0.02\tabularnewline
			\hline 
			$^{34}\text{Cl}\rightarrow^{34}\!\!\text{S}$ & -0.17 & -0.21 & -0.16 & -0.06 & -0.04 & 0.08 & 0.10 & 0.07 & 0.03 & 0.02\tabularnewline
			\hline 
			$^{38m}\text{K}\rightarrow^{38}\!\!\text{Ar}$ & -0.15 & -0.42 & -0.15 & -0.07 & -0.04 & 0.06 & 0.16 & 0.06 & 0.03 & 0.01\tabularnewline
			\hline 
			$^{42}\text{Sc}\rightarrow^{42}\!\!\text{Ca}$ & -0.15 & -0.17 & -0.09 & -0.07 & -0.04 & 0.05 & 0.06 & 0.03 & 0.02 & 0.01\tabularnewline
			\hline 
			$^{46}\text{V}\rightarrow^{46}\!\!\text{Ti}$ & -0.12 & -0.11 & -0.08 & / & -0.04 & 0.03 & 0.03 & 0.02 & / & 0.01\tabularnewline
			\hline 
			$^{50}\text{Mn}\rightarrow^{50}\!\!\text{Cr}$ & -0.12 & -0.09 & -0.06 & / & -0.04 & 0.03 & 0.02 & 0.02 & / & 0.01\tabularnewline
			\hline 
			$^{54}\text{Co}\rightarrow^{54}\!\!\text{Fe}$ & -0.13 & -0.10 & -0.07 & -0.05 & -0.05 & 0.03 & 0.02 & 0.02 & 0.01 & 0.01\tabularnewline
			\hline 
		\end{tabular}
		\par\end{centering}
	\caption{\label{tab:finalB}Estimation of $\Delta M_B^{(1)}$ and $|\Delta M_B^{(1)}/(AR^2/2)|$ from different models.}
\end{table*}

{\bf\boldmath Isovector monopole dominance} -- 
An even more straightforward relation between $\Delta M^{(1)}_{A,B}$ and
$\delta_{\text{C}}$ can be established by invoking the concept of isovector monopole dominance~\cite{AUERBACH1983273,Auerbach:2008ut}, which states that the sum over reduced matrix elements of the isovector monopole operator is largely saturated by the contribution from the giant isovector
monopole states (IVMS) which we denote as $|M;T,T_{z}\rangle$, with energies $E_{M,T}$.
Furthermore, it is argued that the difference between the reduced
matrix elements at different isospin channels of $|M;T\rangle$ are
of the order $(N-Z)/A\ll1$. Hence, in this approximation scheme all matrix elements are equal,
$\langle M;T||V_{C}^{(1)}||g;1\rangle\equiv u$ for $T=0,1,2$. 
From Eq.\eqref{eq:DeltaC} it appears that for $\delta_{\text{C}}$ to be non zero, a splitting between the IVMS energies in different isospin channels $E_{M,0}$, $E_{M,1}$,
$E_{M,2}$ must be introduced. This splitting comes about from the symmetry
potential with the result from Ref.~\cite{Auerbach:2008ut},
\begin{equation}
E_{M,T}-E_{g,1}=\xi\omega[1+(T^2+T-4)\kappa/2],~~T=0,1,2
\end{equation}
with $\kappa\equiv2V_{1}/(\xi\omega A)$, $V_{1}$ the strength of the symmetry potential, $\omega$ the harmonic oscillator frequency,
and $\xi$ a model parameter describing the IVMS strength. With these ingredients we obtain:
\begin{equation}
\delta_{\text{C}}\approx\frac{\kappa(4-13\kappa+12\kappa^{2}-\kappa^{3})}{(1-2\kappa)^{2}(1-\kappa^{2})^{2}}\frac{u^{2}}{\xi^{2}\omega^{2}},
\end{equation}
we see that it is suppressed by the small energy splitting parameter $\kappa$. 
The same treatment applies to $\Delta M^{(1)}_{A,B}$; they are all proportional
to the same unknown reduced matrix element $u^{2}$, and could be connected to $\delta_\text{C}$ as:
\begin{eqnarray}
\delta_{\text{C}}&\approx&-\frac{Ze^{2}}{8\pi R_C^{3}}\frac{\kappa(4-13\kappa+12\kappa^{2}-\kappa^{3})}{(\kappa^2-4\kappa+2)(1-2\kappa)(1-\kappa^2)}\frac{1}{\xi\omega}\Delta M^{(1)}_A\nonumber\\
&\approx&-\frac{Ze^{2}}{8\pi R_C^{3}}\frac{(4-13\kappa+12\kappa^{2}-\kappa^{3})}{2\kappa(1-2\kappa)(1-\kappa^2)}\frac{1}{\xi\omega}\Delta M^{(1)}_B~,\label{eq:proportion}
\end{eqnarray}
where $u^2$ now drops out.
Hence we have obtained a direct relation between $\delta_{\text{C}}$
and $\Delta M^{(1)}_{A,B}$, with a proportionality constant bearing a residual model dependence. We notice that $\Delta M_A^{(1)}$ is not suppressed by $\kappa$, so its sensitivity to $\delta_\text{C}$ is enhanced by $1/\kappa$; on the other hand $\Delta M_B^{(1)}$ is suppressed by $\kappa^2$ so it requires a much higher experimental precision to observe a deviation from zero. Furthermore, the ratio between $\Delta M_{A,B}^{(1)}$ depends only on $\kappa$,
so a simultaneous measurement of the two may pin down $\kappa$, which further solidifies their relation to $\delta_\text{C}$.

{\bf\boldmath Targeted experimental precision} -- Following the strategy outlined above, we devise the experimental precision required for the quantities $\Delta M_{A,B}^{(1)}$, which would allow to address the reliability of the estimates of $\delta_{\text{C}}$ and its uncertainty in a less model-dependent way. 
First, to fix the proportionality constant, we take:
\begin{equation}
Z\approx{A}/{2},\:\:R_C\approx\sqrt{{5}/{3}}\times 1.1\text{fm}\times A^{1/3},
\end{equation}
with the standard expectation for the nuclear RMS radius, $R\approx 1.1\text{fm}\times A^{1/3}$, related to the radius of a nucleus as a uniform sphere by $R^2=(3/5)R_C^2$. We take further parameters from Ref.~\cite{Auerbach:2008ut},  
\begin{equation}
V_{1}\approx100\text{MeV},\:\:\omega\approx41\text{MeV}\times A^{-1/3},\:\:\xi\approx3.\label{eq:parameters}
\end{equation}
More recent discussions of these parameters supporting the above choices can be found in Refs.~\cite{Loc:2018pgm,Auerbach:2021jyt}.
Next, we may, e.g., take the estimates of
$\delta_{\text{C}}$ available in the literature and substitute them into the first line of Eq.~\eqref{eq:proportion}. This returns an estimate of
the size of $\Delta M^{(1)}_A$, which informs, how precise the measurement of this quantity should be to discriminate the model dependence of $\delta_{\text{C}}$.

Restricting ourselves to superallowed decays with $T_{z,i}=0$ and $T_{z,f}=+1$ and requiring the daughter
nucleus to be (observationally) stable, we study the transitions with $26\leq A\leq 54$. 
We take $\delta_\text{C}$ as calculated in the nuclear shell model
with the Woods-Saxon (WS) potential~\cite{Hardy:2020qwl}, the density functional theory (DFT)~\cite{Satula:2016hbs}, the Hartree-Fock (HF) calculation~\cite{Ormand:1995df}, the random phase approximation (RPA) with PKO1 parameterization~\cite{Liang:2009pf}, as well as the ``miscroscopic'' model of Ref.\cite{Auerbach:2008ut,Auerbachwrong} which gives $\delta_\text{C}\approx 2\times 18.0\times 10^{-7}A^{5/3}$.
The estimated size of $\Delta M^{(1)}_A$ indicates the targeted absolute precision in the measurements of $\langle f|M_{+1}^{(1)}|i\rangle$ and $\langle f|M_{0}^{(1)}|f\rangle$. The latter implies subtracting two large terms, 
$NR_{n,f}^{2}/2$ and $ZR_{p,f}^{2}/2$, each
of the typical size $AR^{2}/4$. Therefore, we may use the ratio $\Delta M^{(1)}_A/(AR^{2}/4)$
as an estimate of the precision of the RMS radii of the nuclear neutron and proton distributions required to probe the ISB effects.

The results of our numerical analysis are summarized in Table~\ref{tab:final}. We find that most models predict a generic size of $\Delta M^{(1)}_A\sim 1\text{fm}^2$, with a precision level $(1-3)\%$ needed for the $R_{p,f}^2$ and $R_{n,f}^2$ measurements in order to probe the isospin mixing effect, i.e. start seeing a deviation of $\Delta M^{(1)}_A$ from zero. If it turns out that a non-zero $\Delta M^{(1)}_A$ is not observed at this precision, it could indicate that the actual values of $\delta_\text{C}$ are smaller than most existing model predictions, as  suggested in~\cite{Miller:2009cg,Condren:2022dji}.
The model predictions for $\Delta M^{(1)}_A$ span over an order of magnitude for $^{38m}$K$\to\,^{38}$Ar, and half that range for $^{34}$Cl$\to\,^{34}$S and $^{42}$Sc$\to\,^{42}$Ca decays, reflecting a similar model dependence in $\delta_\text{C}$ in these channels. Hence, an experimental study of $\Delta M^{(1)}_{A}$ for these systems even at a moderate precision will shed light on the model dependence of  $\delta_\text{C}$.
An analogous analysis for $\Delta M_B^{(1)}$ is summarized in Table~\ref{tab:finalB}; following Eq.\eqref{eq:DeltaMB1}, we use $\Delta M_B^{(1)}/(AR^2/2)$ as a measure of the precision goal. We observe that, due to the $\kappa^2$-suppression, a much higher precision (0.01-0.1)\% is required to probe $\delta_\text{C}$ experimentally through $\Delta M_B^{(1)}$.

{\bf\boldmath Discussion of the experimental feasibility} -- To constrain $\Delta M_A^{(1)}$ we need $R_\text{Ch}^2$ and $R_\text{NW}^2$ for the stable nucleus, as well as $R_\text{CW}^2$. Considering $A=38$ where the spread in model predictions is as large as an order of magnitude~\ref{eq:deltaMA}, even a 10\% precision of these radii allows to discriminate between models. The typical $R_\text{Ch}^2$ precision is per mille or better. $R_\text{NW}^2$ remains to be measured in fixed-target electron-nucleus scattering experiments.  The recent PREX-2 and CREX experiments measured $R_\text{NW}$ of $^{208}$Pb and $^{48}$Ca to 1.4\%~\cite{PREX:2021umo} and 0.95\%~\cite{CREX:2022kgg} respectively. The future P2 experiment~\cite{Becker:2018ggl} plans to further improve this precision, and a measurement of $R_\text{NW}$ of $^{12}$C to 0.5\% or better is feasible~\cite{Koshchii:2020qkr}. This allows us to  believe that our desired precision goal for $R_\text{NW}$ for the stable $T=1$ nucleus is achievable with the available experimental technique.

To extract $R_\text{CW}^2$ one would need to measure recoil effects in the superallowed beta decay itself. Recoil effects in beta decays have been measured before to extract the $\beta-\nu$ correlation, e.g. at TRIUMF and CERN~\cite{Gorelov:2004hv,VanGorp:2014cpa,Araujo-Escalona:2019ijw}.
The main challenge of this approach is the smallness of the $Q$-value of the decay, e.g. 
$Q_\text{EC}\approx 6.04$~MeV~\cite{Hardy:2020qwl} for
$^{38m}$K$\rightarrow$$^{38}$Ar. Inserting this into Eq.\eqref{eq:fbart} gives $Q_\text{EC}^2 R_\text{CW}^2/6\approx 0.25\%$, implying a $\sim$0.02\% measurement of the beta-decay spectrum for the $\sim$10\% precision goal for $R_\text{CW}^2$, which is quite challenging and may be taken as a long-term goal. All other SM corrections to the decay spectrum, such as the shape factor $C(Z,W)$ (see, e.g. Ref.~\cite{Hayen:2017pwg} for a detailed discussion), need to be under control. Neutrino-nucleus scattering $ \phi_f \nu\rightarrow \phi_i \ell$ which probes exactly the same charged weak form factor $\bar{f}_+(t)$ offers another possibility. Since the neutrino beam energies are typically much larger (e.g. $E_\nu\approx 0.6$~GeV at T2K and Super-K~\cite{T2K:2016jor}, and even higher at DUNE~\cite{DUNE:2015lol}), $Q^2 R_\text{CW}^2/6$ could be of the order unity or larger and a 10\% determination of $R_\text{CW}^2$ is more feasible. Experimental challenges, such as the detection of the final-state nucleus and the determination of the momentum exchange, must be properly addressed but are beyond the scope of this paper.

In turn, as an estimate we can relate $R_\text{CW}$ to the charge radii across the isotriplet in the isospin-symmetric limit,
\begin{eqnarray}
R_\text{CW}^2&=&R_{\text{Ch},1}^2+Z_0(R_{\text{Ch},0}^2-R_{\text{Ch},1}^2)\nonumber\\
&=&R_{\text{Ch},1}^2+Z_{-1}(R_{\text{Ch},-1}^2-R_{\text{Ch},1}^2)/2~.\label{eq:RCWrelation}
\end{eqnarray} 
It states that $R_\text{CW}$ can be predicted, modulo small ISB corrections, if two out of three charge radii in an isotriplet are known.
For $A=38$, we take the experimental values of $R_\text{Ch}({}^{38}\text{Ar})=3.4028(19)$~fm~\cite{Angeli:2013epw} and $R_\text{Ch}({}^{38}\text{Ca})=3.467(1)$~fm~\cite{C38chargeradius} to obtain
$R_\text{CW}^2= 15.99(28)$~fm$^2$, which we used above for the estimation of the size of the recoil effects. 

Next we turn to $\Delta M_B^{(1)}$. It can readily be computed for $A=38$ isotriplet where all charge radii are known: in addition to the other two mentioned above, one has $R_\text{Ch}({}^{38m}K)=3.437(4)$~fm~\cite{Bissell:2014vva}. 
We obtain $\Delta M_B^{(1)}(A=38)=-0.00(12)(52)(14)$~fm$^2$, with the three respective uncertainties corresponding to those of the charge radii of the $T_z=-1,0,1$ member of the isotriplet. 
Comparing this result with the entry in Table~\ref{tab:finalB} we see that a future improvement of the $^{38m}$K charge radius precision 
by a factor $1.5-2$ will already have practical 
implications for $\delta_\text{C}$. This may serve as the first experimental goal along this direction, feasible in the near future. Other isotriplets may require a higher precision of the charge radii to probe $\delta_\text{C}$, which may serve as a long-term goal.

Guided by the example of $\Delta M_B^{(1)}(A=38)$, we can turn the argument around: taking $\Delta M_B^{(1)}=0$ in Eq.\eqref{eq:DeltaMB1}, one could predict the third charge radius to a 0.1\% accuracy if the other two are known. It is well-known that one of the major problems of charge radius measurements is the lack of absolute charge radii, from which charge radii of other unstable isotopes could be determined through isotopic shift. Requiring $\Delta M_B^{(1)}=0$ thus provides the absolute scale for charge radii of unknown isotopes at the level of $0.1$\%, which could be very useful. This serves as an extra motivation for new measurements of charge radii across various isotriplets, beyond constraining $\delta_\text{C}$.

{\bf\boldmath Summary} -- We propose the possibility to constrain the ISB correction $\delta_\text{C}$ to the rates of superallowed nuclear beta decays by experimental data. 
Our formalism combines that proposed by Miller and Schwenk~\cite{Miller:2008my,Miller:2009cg,Miller:2020xhy} and by Auerbach~\cite{Auerbach:2008ut}, but enlarges the scope of both. In view of a significant model spread of the ISB correction calculations, we opt for constraints from two thoroughly constructed combinations of measurable quantities, $\Delta M^{(1)}_{A}$ and $\Delta M^{(1)}_{B}$.
The former is 
related to nuclear weak radii, while the latter to nuclear charge radii. The information on some nuclear charge radii has been used by Hardy and Towner in the past~\cite{Hardy:2014qxa}. Under the assumption that ISB is predominantly isovector, (to be tested in future work) we unambiguously connect $\delta_\text{C}$ to the charge radii differences across the superallowed isotriplet. 
The inclusion of the weak nuclear radii, both in the charged current (via the measurement of the nuclear recoil in the superallowed transition) and in the neutral current (via the measurement of the neutron skin of the stable daughter nucleus with PVES) is new. In a simplified picture with the IVMS dominance, $\Delta M^{(1)}_{A,B}$ and $\delta_\text{C}$
are all unambiguously interconnected. We individuated transitions in the superallowed decay chart where a measurement $\Delta M_A^{(1)}$
at even a moderate, few percent precision, could already discriminate between models used to compute $\delta_\text{C}$; $\Delta M_B^{(1)}$ requires higher precision but partial information already exists.
Moreover, this study suggests affinities, never attenuated earlier, between experimental programs and communities in physics of rare isotopes,  electron scattering and nuclear astrophysics. 

\begin{acknowledgments}
	
	{\bf\boldmath Acknowledgments} -- We thank Minh-Loc Bui, Gerald Miller, Kei Minamisono, Sonia Bacca and Xavier Roca-Maza for useful conversations. The work of C.Y.S. is supported in
	part by the Deutsche Forschungsgemeinschaft (DFG, German Research
	Foundation), by the NSFC through the funds provided to the Sino-German Collaborative Research Center TRR110 ``Symmetries and the Emergence of Structure in QCD'' (DFG Project-ID 196253076 - TRR 110, NSFC Grant No. 12070131001), by the U.S. Department of Energy (DOE), Office of Science, Office of Nuclear Physics, under the FRIB Theory Alliance award DE-SC0013617, and by the DOE grant DE-FG02-97ER41014. The work of M.G. is supported in part by EU Horizon 2020 research and innovation programme, STRONG-2020 project
	under grant agreement No 824093, and by the Deutsche Forschungsgemeinschaft (DFG) under the grant agreement GO 2604/3-1. 
	
\end{acknowledgments}

\bibliography{deltaC_ref}

\end{document}